\def \SAIT #1 #2 {{\em Mem.\ Soc.\ Astron.\ It.\/} {\bf #1}, #2}
\def \MESS #1 #2 {{\em The Messenger\/} {\bf #1}, #2}
\def \ASTRNACH #1 #2 {{\em Astron. Nach.\/} {\bf #1}, #2}
\def \AAP #1 #2 {{\em Astron. Astrophys.\/} {\bf #1}, #2}
\def \AAL #1 #2 {{\em Astron. Astrophys. Lett.\/} {\bf #1}, L#2}
\def \AAR #1 #2 {{\em Astron. Astrophys. Rev.\/} {\bf #1}, #2}
\def \AAS #1 #2 {{\em Astron. Astrophys. Suppl. Ser.\/} {\bf #1}, #2}
\def \AJ #1 #2 {{\em Astron. J.\/} {\bf #1}, #2}
\def \ANNREV #1 #2 {{\em Ann. Rev. Astron. Astrophys.\/} {\bf #1}, #2}
\def \APJ #1 #2 {{\em Astrophys. J.\/} {\bf #1}, #2}
\def \APJL #1 #2 {{\em Astrophys. J. Lett.\/} {\bf #1}, L#2}
\def \APJS #1 #2 {{\em Astrophys. J. Suppl.\/} {\bf #1}, #2}
\def \APSS #1 #2 {{\em Astrophys. Space Sci.\/} {\bf #1}, #2}
\def \ASR #1 #2 {{\em Adv. Space Res.\/} {\bf #1}, #2}
\def \BAIC #1 #2 {{\em Bull. Astron. Inst. Czechosl.\/} {\bf #1}, #2}
\def \JSQRT #1 #2 {{\em J. Quant. Spectrosc. Radiat. Transfer\/} {\bf #1}, #2}
\def \MN #1 #2 {{\em Mon. Not. R. Astr. Soc.\/} {\bf #1}, #2}
\def \MEM #1 #2 {{\em Mem. R. Astr. Soc.\/} {\bf #1}, #2}
\def \PLR #1 #2 {{\em Phys. Lett. Rev.\/} {\bf #1}, #2}
\def \PASJ #1 #2 {{\em Publ. Astron. Soc. Japan\/} {\bf #1}, #2}
\def \PASP #1 #2 {{\em Publ. Astr. Soc. Pacific\/} {\bf #1}, #2}
\def \NAT #1 #2 {{\em Nature\/} {\bf #1}, #2}

\documentstyle{memsait}
\input epsf.sty
\begin{opening}
\title{EUVE-RXTE SIMULTANEOUS OBSERVATIONS OF NGC~4051: VARIABILITY AND SPECTRUM} 
\author{I. Cagnoni$^{1,2}$, A. Fruscione$^2$, I. M. McHardy$^3$, I. E. Papadakis$^4$}
\institute{$^1$SISSA-ISAS, Trieste, Italy\\
$^2$CfA, Cambridge, USA\\
$^3$University of Southampton, Southampton, UK\\
$^4$University of Crete, Crete}
\date{} 
\end{opening}

\begin{document}

\oddpagefooter{}{}{} 
\evenpagefooter{}{}{} 
\ 
\bigskip

\begin{abstract}
We present timing and spectral analysis of the data collected by the
Extreme Ultraviolet Explorer Satellite ({\it EUVE}) for the Seyfert 1 galaxy
NGC 4051 during 1996.  NGC 4051 was observed twice in May 1996 and
again in December 1996 for a total of more than 200 ks.  The
observations were always simultaneous with hard X-ray observations
conducted with the Rossi X-Ray Timing Explorer ({\it RXTE}).  The EUVE light curves
are extremely variable during each observation, with the maximum
variability of more than a factor of 15 from peak to
minimum.  We detected signal in the EUVE spectrograph in the 75-100 \AA \/
range which is well fitted by absorbed power law models.  
We illustrate the results of our spectral and detailed power
spectrum analysis performed on {\it EUVE} data and  the comparison with
{\it RXTE} lightcurves and discuss the constraint we can place on the 
mass of the central object and on the size of the emitting region.
\end{abstract}

\section{Introduction}
 
NGC~4051 (z=0.0023) is a low-luminosity narrow line Seyfert~1 galaxy which displays
strong and rapid variability in the X-rays (e.g. Papadakis \& Lawrence 1995)
while its optical flux has not been seen to vary within 1 per cent even when the X-ray
flux varied by a factor of 2 (Done et al. 1988).
 Given the presence of a very strong soft X-ray excess, the close
distance and the low Galactic hydrogen column, 
N$_H = 1.31\times 10^{20}$~$cm^{-2}$ (Elvis, Lockman \& Wilkes 1989), 
NGC~4051 was expected to be a strong source of extreme ultraviolet (EUV) radiation 
(58--740 \AA\ or 0.02--0.21 keV) and indeed 
it was seen ($3.6 \sigma$ detection, Fruscione 1996) by the 
{\it Extreme Ultraviolet Explorer
Satellite}  ({\it EUVE}) during the all-sky survey.

We obtained simultaneous observations of NCG~4051 with the 
{\it Rossi X-ray Timing Explorer} ({\it RXTE}) and with the {\it EUVE} 
satellite to investigate if the source is variable in the EUV and if and how 
this energy band is related to the X-rays.

\section{Observations}

NGC~4051 was observed with the Deep Survey-Spectrometer (DS/S) 
onboard the {\it EUVE} satellite twice in May 1996 and once
in December 1996 for a total of approximately 200 ks. 
The DS/S  configuration (e.g. Welsh et al. 1990)
allows simultaneous imaging and spectroscopy in the 58--740 \AA\ 
(0.02--0.21 keV) region with a spatial resolution of $\sim 1^{\prime}$ and a 
spectral resolution of $\lambda/\Delta \lambda \sim 200$ at the short 
wavelengths.

In May 1996 {\it RXTE} was observing NGC~4051 with several short observations 
(typical duration $\sim 1$ ks),
while from December 13 to December 16 {\it RXTE} performed a long look
which overlapped for $\sim 120$ ks with {\it EUVE}.
Only {\it RXTE} data from the Proportional Counter Array (PCA),
which covers the 2-60 keV band, were used.

\section{{\it EUVE}: energy spectrum, lighcurves and power spectrum}

\begin{figure}
\epsfysize=5.3cm 
\hspace{3.5cm}\epsfbox{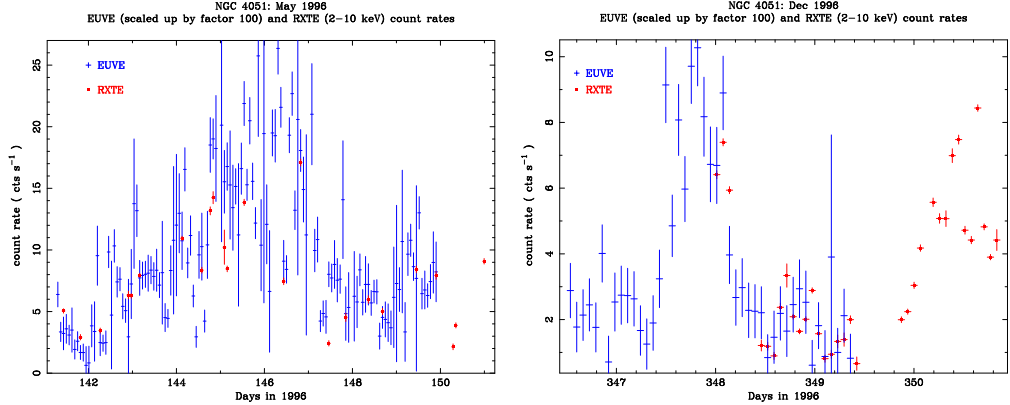} 
\caption[h]{May (left) and December (right) 1996: 5544 s binned {\it EUVE} (scaled by
a factor 100) and {\it RXTE} (2-10 keV) light curves of NGC~4051 
(the errors are $1 \sigma$)}
\end{figure}

NGC~4051 was detected by {\it EUVE} only in the region below $\sim 100$ \AA\ because 
the interstellar medium severely
attenuates the EUV spectrum at longer wavelengths; despite the
long ($\sim 203$ ks) total effective exposure, the signal to noise ratio
(SNR) achieved in the spectrum is very low ($\sim 1.5$ in the range 80-90 \AA).
We modeled the EUV spectrum of NGC~4051 over the 70-100 \AA \/ range with an
absorbed power law of the form

\[F(E)=k E^{-\alpha} \exp{(-{\Sigma}_{x} N(X) {\sigma}_X )}\]

\noindent where $k$ is a normalization factor, $\alpha$ is the energy index and
the absorption is characterized by  a column density N(X) and an
absorption cross section $\sigma_{X}$ for each element.
We fixed the Galactic hydrogen column density at $N_H =1.31 \times
10^{20}$~cm$^{-2}$ (Elvis et al. 1989) and the ratios He{\sc i}/H{\sc i}=0.1 
and He{\sc ii}/H{\sc i}=0.01.
The best fit values are: $\alpha = 0.9^{+1.8}_{-1.9}$ and $k=70^{+13}_{-9} \mu
Jy$ (at 0.155 keV).
Further information regarding the {\it EUVE} 
spectrum can be found in Fruscione et al. 1998.\\

Thanks to the $\sim 10$ times larger effective area of the DS instrument, a much higher
SNR can be achieved in the lightcurves: the average SNR in each {\it EUVE} orbit
(5544 s on average) is about 6.5.
{\it EUVE} lightcurves binned over 5544 s  are presented, together with {\it RXTE}
data, in
Figure~1; {\it EUVE} count rates were multiplied by a factor of 100 for presentation
purposes.
Both during May 20-28 and December 11-14, NGC~4051 was very variable showing changes by
a factor $\sim 15$ from peak to minimum; in December the source was in a lower state
(the average count rate in May was CR$= 0.0937 \pm 0.0011$ counts $s^{-1}$ and in
December CR$=0.0346 \pm 0.0012$ counts $s^{-1}$) but showed similar variability.\\

We performed a power spectrum analysis 
on the lighcurves binned over 500 s (for details see Fruscione et al. 1998);
 the averaged and binned results are presented in Figure~2.
We fitted the logarithm of the power spectrum with a model of the form
\begin{equation}
P_{mdl}(\nu)=\frac{A}{[(2\pi\nu)^{2}+f_{b}^{2}]^{a}}+C.
\end{equation}
\noindent (which is a power law with slope $2a$ for frequencies $ >> f_b/2\pi$ and flattens at
frequencies $<< f_b/2\pi$; the constant $C$ represents the power level due to the
Poisson counting statistics) convolved with {\it EUVE} window function.
The peaks in Figure~2 are due to the satellite orbital period and one of its harmonics.

Both May and December 1996 power spectra for NGC~4051 are well fitted by this model; 
the best fit parameters for the two power spectra are consistent within the 68\%
confidence regions.
The slope of the power spectrum is steeper than 1.6, and the spectrum flattens at
frequencies lower than $f_b/2\pi = [4.9 \times 10^{-5} - 1.7 \times 10^{-5}]$ Hz.

\begin{figure}
\epsfysize=6cm 
\hspace{3.5cm}\epsfbox{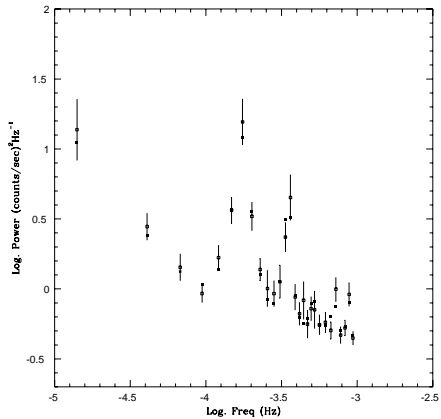} 
\caption[h]{Power spectrum obtained from May 1996 lightcurve binned over 500 s.
The strong peaks correspond to {\it EUVE} orbital period and one of its harmonics.}
\end{figure}

\section{Comparison of {\it EUVE} and {\it RXTE} lightcurves}

A comparison of {\it EUVE} and {\it RXTE} lightcurves, binned over one average {\it
EUVE} orbit (5544 s), is presented in Figure~1, where
{\it EUVE} data were scaled up of a factor of 100 for presentation purposes.
The {\it RXTE} lightcurves were extracted in the 2-10 keV region where the count
rate varies from 3 to 17.4 counts $s^{-1}$.
A striking correlation between the two bands is clearly visible.
We performed a cross-correlation analysis with the Discrete Correlation Function (DCF)
of Edelson \& Krolik 1988, between the {\it EUVE} band and the 4-10 keV {\it RXTE}
band, excluding the 5-7 keV interval, to avoid both the iron line and the warm absorber.
The DCF confirms that the best estimate of the lag is zero, although it is possible
that the EUV band may lead the X-rays by $\sim 0.2$ days.
(Further details can be found in Uttley et al. 1998)

\section{Discussion}

This first deep {\it EUVE} observation of NGC~4051 allowed a detailed power spectrum
analysis: the QPO found in EXOSAT data cannot be confirmed since its frequency is very
similar to {\it EUVE} orbit.
If the break, found in the best fit of the power spectrum, were related to
thermal instabilities in the accretion disk (Nowak 1994) and if we assume 
that the EUV emission comes from 7R$_S$ and keplerian orbit, we find that 
the mass of the central black hole lies in the range 
$2 - 8 \times 10^{7}$ M$_{\odot}$.

From the comparison of {\it EUVE} and {\it RXTE} lightcurves it is clear that 
there is a tight correlation between the EUV and the continuum X-ray emission;
no lag was found down to 0.2 days and it could be further 
reduced down to 1 ks (Uttley et al. 1998).
Assuming that the observed lag corresponds to the time interval needed to 
upscatter EUV photons to X-ray energies, we can constrain the size of the 
Comptonising region
for a $10^6$ M$_{\odot}$ black hole to $\sim 400$ R$_S$ for 0.2 day lay or 
to $\sim  20$ R$_S$ for 1 ks lag (see details in Uttley et al. 1998).

\acknowledgements

This work was supported by AXAF Science Center NASA Contract NAS 8-39073 and by NASA
grant NAG 5-3174 and NAG 5-3191.

\end{document}